\documentclass{amsproc}

\usepackage{amsmath}
\usepackage{amssymb}
\usepackage{graphicx}

\theoremstyle{definition}

\theoremstyle{remark}

\numberwithin{equation}{section}

\def\erf{{\rm erf}}
\def\rv{{\bf r}}

\def\Rv{{\bf R}}

\def\beq{\begin{equation}}
\def\eeq{\end{equation}}

\begin{document}

\title{Model hamiltonians in density functional theory}

\author{Paola Gori-Giorgi}
\address{Laboratoire de Chimie Th\'eorique, CNRS,
Universit\'e Pierre et Marie Curie, 4 Place Jussieu,
F-75252 Paris, France}
\email{gori@lct.jussieu.fr}
\author{Julien Toulouse}
%%    \thanks will become a 1st page footnote.
\address{Cornell Theory Center,
Frank H.T. Rhodes Hall,
Cornell University,
Ithaca, New York, 14853-3801, USA.
}
\email{toulouse@cornell.edu}
% %   Information for second author
\author{Andreas Savin}
\address{Laboratoire de Chimie Th\'eorique, CNRS,
Universit\'e Pierre et Marie Curie, 4 Place Jussieu,
F-75252 Paris, France}
\email{savin@lct.jussieu.fr}

%    General info
%\subjclass{Primary 54C40, 14E20; Secondary 46E25, 20C20}
%\date{January 1, 1994 and, in revised form, June 22, 1994.}



\begin{abstract}
The formalism of Kohn and Sham uses a specific (model) hamiltonian which highly simplifies the many-electron problem to that of noninteracting fermions. The theorem of Hohenberg and Kohn tells us that, for a given ground state density, this hamiltonian is unique. In principle, this density can be chosen as that of the real, interacting system. To obtain the energy, or other properties of the real system, approximations are needed. Working with non
interacting fermions is an important simplification, but it may be
easier to produce approximations with different choices of the model
hamiltonian. The feature that the exact density is (ideally)
reproduced can be kept in the newly defined fictitious systems. Using
model hamiltonians having the same form as the physical one, that is,
being built of one- and two-body operators, allows to approach the
physical hamiltonian arbitrarily close, and thus a systematic reduction
of the approximations. 
\end{abstract}

\maketitle

\section{Introduction}
\subsection{General context}
The knowledge from first principles of the electronic structure
of atoms, molecules and solids is 
contained in the $N$-electron Schr\"odinger equation that,
within the Born-Oppenheimer approximation (i.e., at fixed nuclei
positions $\Rv_{\alpha}$), reads
\beq
\hat{H}(\rv_1,...,\rv_N)\,\Psi(\rv_1\sigma_1,...,\rv_N\sigma_N)=
E\,\Psi(\rv_1\sigma_1,...,\rv_N\sigma_N),
\label{eq_Schro}
\eeq
with
\begin{eqnarray}
\hat{H} & =& \hat{T}+\hat{W}_{ee}+\hat{V}_{ne}, \\
\hat{T} & = & -\frac{1}{2}\sum_{i=1}^N\nabla_{\rv_i}^2, 
\label{eq_T} \\
\hat{W}_{ee} & = & \frac{1}{2}\sum_{i\neq j}^N \frac{1}{|\rv_i-\rv_j|}
\equiv \frac{1}{2}\sum_{i\neq j}^N w_{ee}(|\rv_i-\rv_j|), 
\label{eq_vee} \\
\hat{V}_{ne} & = & -\sum_{i=1}^N\left(\sum_{\alpha=1}^M\frac{Z_{\alpha}}{|\Rv_{\alpha}-\rv_i|}\right) 
\equiv\sum_{i=1}^N v_{ne}(\rv_i),
\label{eq_vne}
\end{eqnarray}
where $v_{ne}$ is the external potential due to the $M$ nuclei of charges $Z_{\alpha}$
at positions $\Rv_\alpha$, and 
Hartree atomic units, $\hbar=1$ (reduced Planck's constant) $m=1$ (electron mass),
$a_0=1$ (Bohr radius), $e=1$ (electron charge), have been used. 
Since electrons obey the Fermi-Dirac statistics, $\Psi$ must be antisymmetric under
particle exchange, $\Psi(\rv_1\sigma_1,...,\rv_i\sigma_i,...,\rv_j\sigma_j,...,
\rv_N\sigma_N)=-\Psi(\rv_1\sigma_1,...,\rv_j\sigma_j,...,\rv_i\sigma_i,...,
\rv_N\sigma_N)$, where $\rv$ denotes the three-dimensional electronic position
and $\sigma$ the spin degree of freedom ($\uparrow$ or $\downarrow$). 
In what follows we will be only concerned
with the search for the ground-state energy, i.e., the lowest eigenvalue $E_0$
of Eq.~(\ref{eq_Schro}).

The methods that both chemists and physicists have developed to find 
approximate solutions of Eq.~(\ref{eq_Schro}) can be roughly divided
in two large groups: wave-function methods
(traditional quantum chemistry methods~\cite{pople},
quantum Monte Carlo~\cite{mitas}) and density methods
(density functional theory~\cite{kohnnobel},
and density matrix functional theory~\cite{coleman}, that is somehow
in between the two
groups). Simplistically, wave-function methods 
start from an approximation for the antisymmetric, normalized, 
$N$-electron wave-function, $\Psi_{\rm approx}$, and take advantage of the
variational principle,
\beq
E_0 \le E_0^{\rm approx}=\min_{\Psi_{\rm approx}}\langle \Psi_{\rm approx}|
\hat{H}| \Psi_{\rm approx}\rangle.
\eeq
Since $\Psi_{\rm approx}$ is a $3N$ dimensional 
object, wave-function methods are in general computationally expensive when
the number of particles increases.
\par
Density Functional Theory (DFT) uses the electron density $n(\rv)$ as a basic
variable, 
\beq
n(\rv)=N\sum_{\sigma_1...\sigma_N}\int |\Psi(\rv\sigma_1,\rv_2\sigma_2,...,
\rv_N\sigma_N)|^2d\rv_2...d\rv_N,
\label{eq_dens}
\eeq
a much simpler quantity to handle, resulting in a low
computational cost that allows to reach system sizes much
larger than those accessible to wave-function methods.
At given number of electrons $N$, the Hohenberg and Kohn 
theorem~\cite{HK} tells us that 
the ground-state density $n(\rv)$
of Eq.~(\ref{eq_dens}) completely determines (except for an
additive constant) the external potential $v_{ne}(\rv)$
of Eq.~(\ref{eq_vne}) (for the sake of simplicity we only consider physical
hamiltonians with a nondegenerate ground state). Since the kinetic energy operator
of Eq~(\ref{eq_T}) and the electron-electron interaction
of Eq.~(\ref{eq_vee})
are the same for all systems, a \textit{universal} 
functional $F$ of the density $n(\rv)$ can be defined as~\cite{mel}
\beq
F[n;\hat{W}_{ee},\hat{T}]=\min_{\Psi \to n}\langle \Psi|\hat{T}
+\hat{W}_{ee}|\Psi\rangle,
\label{eq_Fmel}
\eeq
where, to keep the connection with what we will do in the next sections,
we have explictly shown 
the dependence of $F$ on the electronic interaction $\hat{W}_{ee}$ and on the
kinetic energy operator $\hat{T}$. The minimum search in Eq.~(\ref{eq_Fmel})
is performed over all antisymmetric wave-functions $\Psi$ that yield the
density $n(\rv)$ (by definition $n(\rv)$ also gives, by integration, 
the number of electrons $N$). The
universal functional $F$ can be also defined as a Legendre transform~\cite{lieb} 
\beq
F[n;\hat{W}_{ee},\hat{T}] =\sup_v
\left\{\min_{\Psi} \langle \Psi |\hat{T}+\hat{W}_{ee}+\hat{V}|\Psi\rangle
-\int n(\rv) v(\rv) d\rv\right\},
\label{eq_Flieb}
\eeq
where, as in the rest of this paper, $\hat{V}$ denotes a local one-body
operator of the form of Eq.~(\ref{eq_vne}) with $v_{ne}(\rv)$ replaced
by $v(\rv)$. If the exact form of the functional $F$ was known, the
variational principle would tell us that
\beq
E_0 = \min_n \left\{F[n;\hat{W}_{ee},\hat{T}]+\int n(\rv)\,v_{ne}(\rv)\,
d\rv\right\}.
\label{eq_En}
\eeq 
In practice, because we rely on approximations for $F$, the energy
estimated by carrying out the minimization in Eq.~(\ref{eq_En}) 
can be lower than the exact $E_0$.
\subsection{Kohn-Sham density functional theory}
\label{sec_KSDFT} 
The Kohn-Sham~\cite{KS} approach to DFT introduces another density
functional $T_s[n]$,
\beq
T_s[n]\equiv F[n;0,\hat{T}]=\sup_v
\left\{\min_{\Phi} \langle \Phi |\hat{T}+\hat{V}|\Phi\rangle
-\int n(\rv) v(\rv) d\rv\right\},
\label{eq_Ts}
\eeq
where, as in the rest of this work, $\Phi$ always denotes the wave-function
of a spin-$\frac{1}{2}$ fermionic system with zero electron-electron 
interaction, i.e., in the majority of cases, a single Slater
determinant (the fundamental idea of Kohn-Sham has been to introduce
the fermionic statistic in the construction of $T_s[n]$). Since
dealing with noninteracting particles is computationally simple, Kohn and
Sham~\cite{KS} proposed to search for that particular noninteracting system 
which has the same ground-state density of the physical one. This defines a
model system which is usually called Kohn-Sham (KS) system. The difference
between  $F[n;\hat{W}_{ee},\hat{T}]$ and $T_s[n]$ defines
the Hartree-exchange-correlation
functional $E_{\rm Hxc}[n]$,
\beq
E_{\rm Hxc}[n]=F[n;\hat{W}_{ee},\hat{T}]-T_s[n]
\eeq
 that needs to be approximated. The functional
derivative, $\delta E_{\rm Hxc}[n]/\delta n(\rv)=v_{\rm Hxc}(\rv)$, determines
the one-body KS potential $v_{\rm KS}(\rv)=v_{ne}(\rv)+v_{\rm Hxc}(\rv)$ that
forces the $N$ noninteracting electrons to have the same density
of the physical system. The KS-DFT relies thus on the assumption that, given  
a physical ground-state density $n(\rv)$, it is possible to find
a noninteracting system which has the \textit{same ground-state density}.
\par
From the functional $E_{\rm Hxc}[n]$ the classical electrostatic 
Hartree term is usually extracted,
\beq
E_{\rm H}[n]=\frac{1}{2}\int d\rv\int d\rv'\frac{n(\rv)n(\rv')}{|\rv-\rv'|},
\eeq
and the remaining, unknown, part is called exchange-correlation energy,
\beq
E_{\rm xc}[n]=F[n;\hat{W}_{ee},\hat{T}]-T_s[n]-E_{\rm H}[n].
\eeq 
The Hartree functional $E_{\rm H}[n]$ describes the electrons as 
if they were classical charge distributions. It is
a simple functional of the density, and yields an important
part of the energy of a many-electron system, but is
nonzero also for a one-electron density. This ``self-interaction part'' of 
$E_{\rm H}[n]$ is
cancelled by the exact $E_{\rm xc}[n]$, but this does not occur for
most of the current approximate functionals, which suffer of the so
called ``self-interaction error''.

The success of KS-DFT is mostly due to the fact that simple approximations 
(local-density approximation and generalized gradient corrections)
for $E_{\rm xc}[n]$ and its functional derivative provide 
practical estimates of thermodynamical, structural and spectroscopic properties
of atoms, molecules and solids. However, with the current approximations,
KS-DFT is still lacking in several 
aspects, in particular it fails to handle near-degeneracy correlation 
effects (rearrangement of electrons within partially filled shells) and to 
recover long-range van der Waals interaction energies. 
The inaccuracy of KS-DFT stems from our lack of knowledge of $E_{\rm xc}[n]$,
 and much effort is put nowadays in finding new
approximations to this term~\cite{science}.  A trend in 
the current research is to construct
{\it implicit} functionals of the density: in particular,
the exchange-correlation functional is divided into exact exchange
\beq
E_x[n]=  
\langle \Phi |\hat{W}_{ee}|\Phi\rangle-
E_{\rm H}[n],
\label{eq_ExKS}
\eeq
and the remaining correlation energy $E_c[n]=E_{\rm xc}[n]-E_x[n]$. The
exact exchange cancels the self-interaction error of $E_{\rm H}[n]$,
and is an implicit functional of $n(\rv)$ through the Slater determinant 
$\Phi$. The corresponding Kohn-Sham potential must be determined via
the optimized effective potential method (OEP)~\cite{OEP},
\beq
E_0=\inf_{v}\left\{\langle\Phi_{v}| \hat{T}+\hat{W}_{ee}+\hat{V}_{ne}
|\Phi_{v}\rangle+E_c[n_{\Phi_v}]\right\},
\label{eq_oepKS}
\eeq
where $\Phi_v$ is the ground state of the noninteracting
 hamiltonian, $\hat{T}+\hat{V}$.
 The construction of
a correlation energy functional $E_c[n]$ to be used with exact exchange
in Eq.~(\ref{eq_oepKS}) is still an open problem.

In this work, we review some basic ideas, results, and open questions
of a different approach: 
instead of trying to approximate 
the KS $E_{\rm xc}[n]$, we change the model system defined 
by Eq.~(\ref{eq_Ts}).
\subsection{Adiabatic connection formula}
\label{sec_adiabatic}
Before discussing the choice of different model hamiltonians,
we report some equations that will be used in the next sections.
An exact formula for the functional $E_{\rm Hxc}[n]$ 
can be obtained via the adiabatic connection formalism~\cite{adiabatic,weitao}:
by varying a real parameter $\lambda$,
the interaction $w^{\lambda}(r_{12})$ between the electrons 
(we have defined $r_{12}=|\rv_1-\rv_2|$
to denote pairwise interactions
that only depend on the electron-electron distance)
is switched on continuously from
zero to $1/r_{12}$, while the density is kept fixed by an external
one-body potential $\hat{V}^{\lambda}$. Each hamiltoninan $\hat{H}^\lambda$ 
along this adiabatic connection has a ground-state wavefunction 
$\Psi^\lambda$ that yields, by construction, the same
density $n(\rv)$ for each $\lambda$. If
$w^{\lambda=0}=0$ and $w^{\lambda = a}= 1/r_{12}$, the KS 
Hartree-exchange-correlation energy is given by~\cite{adiabatic,weitao}
\beq
E_{\rm Hxc}[n]=\int_0^a d\lambda \int_0^\infty dr_{12}\,4\pi\,r_{12}^2 
f^{\lambda}(r_{12})\frac{\partial w^{\lambda}(r_{12})}
{\partial \lambda},
\label{eq_adiaKS}
\eeq
where 
the spherically and system-averaged pair density (APD)
$f^{\lambda}(r_{12})$ is obtained by integrating $|\Psi^\lambda|^2$ 
over all variables but $r_{12}=|\rv_2-\rv_1|$,
\beq
f^\lambda(r_{12}) = \frac{N(N-1)}{2}\sum_{\sigma_1...\sigma_N}
 \int |\Psi^\lambda(\rv_{12},\Rv,\rv_3,...,\rv_N)|^2
\frac{d\Omega_{\rv_{12}}}{4\pi} d\Rv d\rv_3...d\rv_N.
\label{eq_intra}
\eeq
where $\Rv=(\rv_1+\rv_2)/2$. 
\section{Changing the model hamiltonian}
\subsection{General considerations}
\begin{figure}
\includegraphics[width=6.7cm]{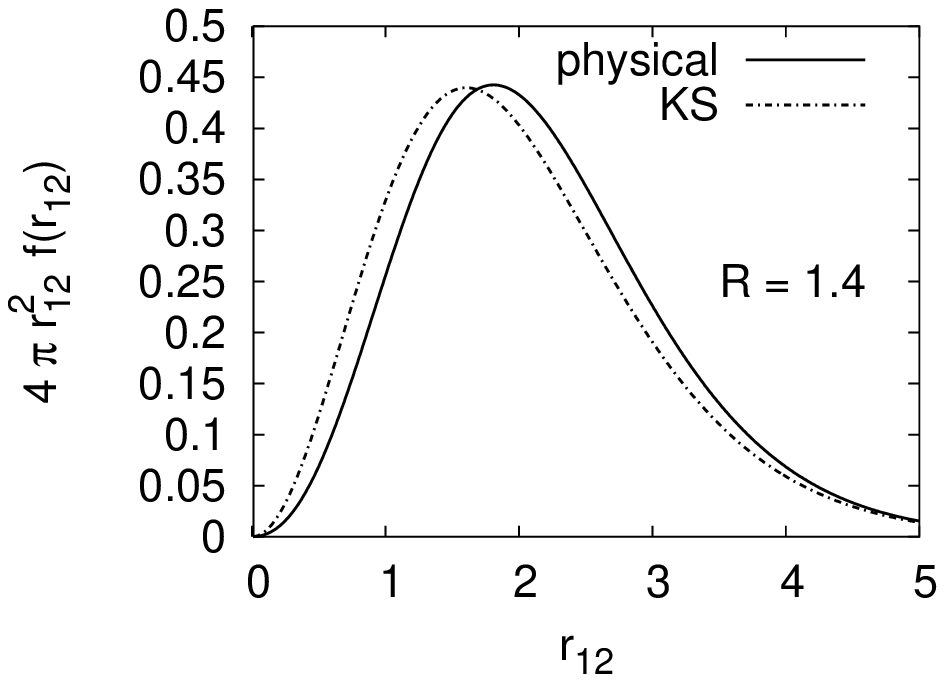} 
\includegraphics[width=6.7cm]{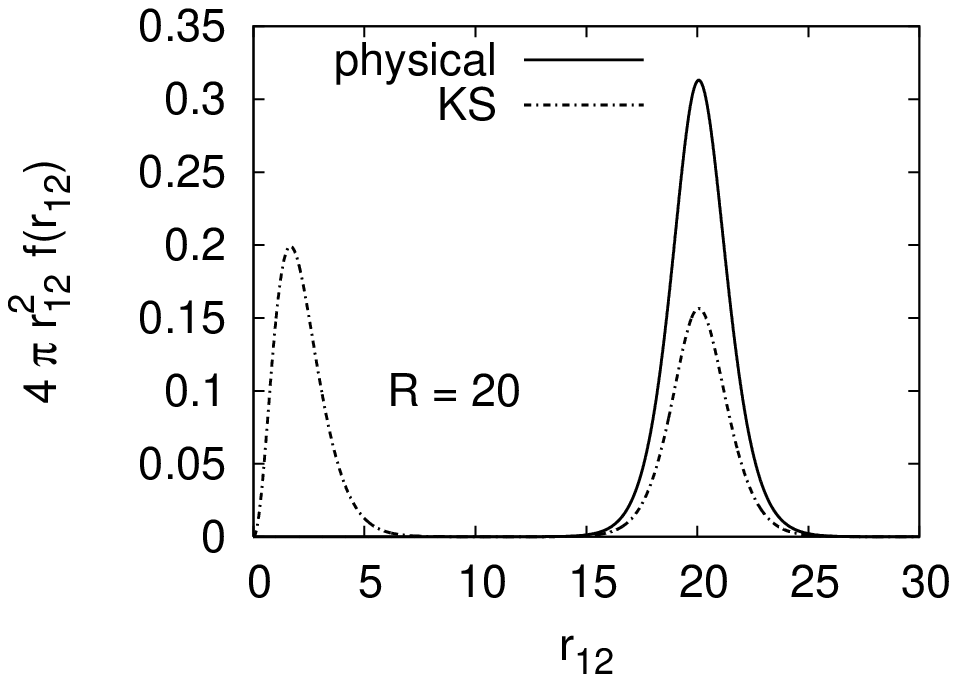} 
\caption{The spherically and system-averaged pair density
$f^\lambda(r_{12})$ of Eq.~(\ref{eq_intra}) 
for the H$_2$ molecule in the KS system ($\lambda=0$) and
in the physical system (Coulombic interaction). The internuclear
equilibrium distance $R=1.4$ and the extreme stretched molecule
at $R=20$ are considered.}
\label{fig_h2}
\end{figure}

Equation~(\ref{eq_Ts}) shows that the KS approach to DFT introduces
a model hamiltonian in which $\hat{W}_{ee}$ is set equal to zero
and the one-body potential, $v_{\rm KS}(\rv)$, is different from 
$v_{ne}(\rv)$ (and is obtained by imposing
the condition that the ground-state electron density of the model hamiltonian
be the same of the physical system).
Solving the noninteracting KS hamiltonian instead of the fully
interacting hamiltonian of Eq.~(\ref{eq_Schro})  
is obviously very practical. The exact ground-state energy of the
physical system can, in principle, be obtained from the KS Slater determinant
$\Phi$ via the functional $E_{\rm xc}[n]$.
However, there are cases in which the restriction that the model
system be noninteracting makes the search for an approximate
$E_{\rm xc}[n]$ seem like a daunting task. 
Consider the simple example of the H$_2$ molecule (two electrons in the field
of two nuclei of unitary charge spaced by a distance $R$).
In Fig.~\ref{fig_h2} we report the quantity
$4 \pi r_{12}^2 f^\lambda(r_{12})$ that enters in Eq.~(\ref{eq_adiaKS})
for the KS system and for the physical system ($\lambda=0$ and $\lambda=a$,
respectively) at two internuclear distances $R$. 
As shown by Eq.~(\ref{eq_adiaKS}), the change in the APD 
$f^\lambda(r_{12})$ when we switch from the
KS system to the physical (fully interacting) system
determines the functional $E_{\rm Hxc}[n]$. If this change is not
drastic, 
we can expect that universal approximations for $E_{\rm Hxc}[n]$
can work relatively well. This is,
e.g., the case of the H$_2$ molecule at the equilibrium distance 
$R=1.4$~a.u., reported in Fig.~\ref{fig_h2}. But when we stretch the molecule
(e.g., in the extreme case $R=20$ considered in the same figure), we see
that the two APD are completely different, which means that the term
$E_{\rm Hxc}[n]$ becomes very important: it has to correct the
very different nature of the KS wavefunction with respect to the physical one.
This effect is completely system-dependent
and it is thus very difficult to include
in a \textit{universal} functional of the density. 
%%%%%%%%%%%%%%%%%%%%%%%%%%%%%%%%%%%%%%%%%%%%%%%%%%%%%%%%%%%%%%%%%%%%
%%%%%%%%%%%%%%%%%%%%%%%%%%%%%%%%%%%%%%%%%%%%%%%%

Cases like this appear when we have near-degenerate levels in the 
Kohn-Sham system
(in the case of the H$_2$ molecule described above, 
the energies of the 
two KS states $\sigma_g$ and $\sigma_u$ become closer and closer as
$R$ increases).
The basic idea reviewed in this paper
 is to remove the constraint that the model system
be noninteracting in order to keep a reasonable resemblance between
the model wavefuntion and the real one. 
%E.g.,
%in the case of the steched H$_2$ molecule, this would correspond in having
%a reference system which is a linear combination of the two states
%$\sigma_g$ and $\sigma_u$.
\subsection{Using a modified electron-electron interaction}
 Obviously, the model system must still be  less
expensive to solve than the physical one. A possible approach
 is to define a density functional for a ``partial'' electron-electron
interaction $w(r_{12})\ge 0$, smaller than the Coulomb interaction
$\forall r_{12}$, $w(r_{12})\le 1/r_{12}$, but different from zero 
(we restrict our choice to pairwise interactions
that only depend on the electron-electron distance).
Following the idea of the adiabatic connection of Sec.~\ref{sec_adiabatic},
we can make our partial interaction depend on a real parameter $\mu$ in such
a way that $w^\mu(r_{12})\to 0$ when $\mu\to 0$ and $w^\mu(r_{12})\to 1/r_{12}$
when $\mu$ tends to some positive value $a$, and define
\beq
F^\mu[n]\equiv F[n;\hat{W}^\mu,\hat{T}] =\sup_v
\left\{\min_{\Psi} \langle \Psi |\hat{T}+\hat{W}^\mu+\hat{V}|\Psi\rangle
-\int n(\rv) v(\rv) d\rv\right\}.
\label{eq_Fmu}
\eeq
With this definition, we see that the density functional $F^\mu[n]$
switches from the Kohn-Sham one of Eq.~(\ref{eq_Ts})
to the physical one of Eq.~(\ref{eq_Flieb}),
\begin{eqnarray}
F^{\mu\to 0}[n] & = & F[n;0,\hat{T}]=T_s[n], \\
F^{\mu\to a}[n] & = & F[n;\hat{W}_{ee},\hat{T}]. 
\end{eqnarray}
In analogy with the KS approach, we can ask that the model system
with interaction $w^\mu(r_{12})$  have the same density of the physical
one. This fixes the external one-body potential 
(that we call $v^\mu(\rv)$) in our model hamiltonian $\hat{H}^\mu$. The ground-state
wavefunction of our model system is denoted $\Psi^\mu$:
it is a multideterminantal wavefunction that must be computed with one
of the standard methods of quantum chemistry (configuration interaction,
coupled cluster, multi-configuration self-consitent field,...). In general,
if $\mu$ is not too large (i.e., if $w^\mu$ is still much smaller
than the full Coulomb interaction), few determinants describe $\Psi^\mu$
quite accurately, so that the computational cost can be kept low.
The difference
$\overline{F}^\mu[n]$ between the physical functional 
$F[n;\hat{W}_{ee},\hat{T}]$ and
the partially-interacting $F^\mu[n]$,
\beq
\overline{F}^\mu[n]=F[n;\hat{W}_{ee},\hat{T}]-F[n;\hat{W}^\mu,\hat{T}],
\label{eq_Fcomp}
\eeq
 is what we need to approximate,
together with its functional derivative that determines 
$v_{ne}(\rv)-v^\mu(\rv)$.
\par
How to choose $w^\mu(r_{12})$? Two points are important to determine the partial interaction:
(i) that we can design reasonable approximations for the corresponding
 $\overline{F}^\mu[n]$, and (ii) that
we can solve adequately the hamiltonian $\hat{H}^\mu$
of the model system. A convenient choice seems to be a long-ranged
interaction, i.e., a $w^\mu(r_{12})$ that behaves as $1/r_{12}$ for large
$r_{12}$, but that is softer than $1/r_{12}$ for small $r_{12}$. The
reasons for this choice are (i) short-range correlation effects seem to be
more transferable from one system to another~\cite{variesucorrSR}, and they  should thus be
more easily described in terms of an 
approximate \textit{universal} density functional; (ii) the
traditional wavefunction methods of quantum chemistry can reasonably
describe $\Psi^\mu$ at a lower cost than the fully interacting
system, because of the smaller interaction and because of the absence of
the electron-electron cusp. For practical reasons (analytic matrix
elements for both Gaussians and plane-waves, i.e. the
most commonly used basis sets in quantum chemistry and solid-state
physics, respectively), a common choice is
\beq
w^\mu(r_{12})=\frac{\erf(\mu\, r_{12})}{r_{12}},
\label{eq_erf}
\eeq
where $\erf(x)$ is the error function (and we thus have $a=\infty$). 
Other possibilities for $w^\mu(r_{12})$ have been also explored: they are all, like 
Eq.~(\ref{eq_erf}), arbitrary, and 
an open question is wether there is a way to determine $w^\mu(r_{12})$ according to
some optimal criteria. The approach resulting from the choice of Eq.~(\ref{eq_erf}) is
the one that we describe more in details in the next Sec.~\ref{sec_lrsr}. Before doing that, for completness
we also briefly introduce a different choice of the model hamiltonian.
\subsection{Adding a nonlocal one-body operator}
Another way to modify the model system is by adding a nonlocal one-body
operator, $\hat{O}_{\rm NL}$, multiplied by a real, positive constant $g$~\cite{rey,claudinetesi},
\beq
F^g[n] \equiv F[n;\hat{T}+g\,\hat{O}_{\rm NL},\hat{W}_{ee}].
\label{eq_nonlo}
\eeq
By setting $\hat{O}_{\rm NL}$ equal to the projector (here reported for closed-shell systems)
onto the virtual Kohn-Sham orbitals $\phi_i(\rv)$ ,
\beq
\hat{O}_{\rm NL}(\rv,\rv')=\delta(\rv-\rv')-\sum_{i=1}^{N/2}\phi_i(\rv)
\phi_i^*(\rv'),
\eeq
the functional $F^g[n]$ of Eq.~(\ref{eq_nonlo}) switches from the physical one at
$g=0$ to the Kohn-Sham one of Eq.~(\ref{eq_Ts}) in the limit $g\to\infty$ (since
$\langle\Psi|\hat{O}_{\rm NL}|\Psi\rangle$ is always $\ge 0$, as $g\to\infty$ the model system minimizes
its energy by making $\langle\Psi|\hat{O}_{\rm NL}|\Psi\rangle=0$, that is, by occupying 
only the KS orbitals).
This choice of the model system is not further detailed here; the interested 
reader may find additional information in 
the literature~\cite{rey,pickett,claudinetesi}. 

\section{Multideterminantal DFT from a long-range-only interaction}
\label{sec_lrsr}
\subsection{The functionals}
\label{sec_functionals}
%%%%%%%%%%% da rimettere?
Following the same steps of Sec.~\ref{sec_adiabatic}, we can write
an exact formula for the
functional $\overline{F}^\mu[n]$ of Eq.~(\ref{eq_Fcomp}) 
in terms of the APD $f^\mu(r_{12})$. It is convenient to use for
the adiabatic connection the same partial interaction
$w^\mu(r_{12})$ of Eq.~(\ref{eq_erf}) chosen to determine the model system,
\beq
\overline{F}^\mu[n]=\int_\mu^\infty 
d\mu' \int_0^\infty dr_{12}\,4\pi\,r_{12}^2 
f^{\mu'}(r_{12})\frac{\partial w^{\mu'}(r_{12})}{\partial \mu'} .
\label{eq_FfromAC}
\eeq
%%%%%%%%%%%%%%%%%%%%%
This formula is identical to  Eq.~(\ref{eq_Fcomp}), except from the
fact that the integration over the coupling constant
starts from the positive value $\mu$ instead than zero. 
\par
As in KS-DFT, the functional $\overline{F}^\mu[n]$ can be divided
into an Hartree term,
\beq
\overline{E}_{\rm H}^\mu[n]= \frac{1}{2}\int d\rv \int 
d\rv' n(\rv)n(\rv')\left[\frac{1}{|\rv-\rv'|}
-w^\mu(|\rv-\rv'|)\right],
\label{eq_EHcomp}
\eeq
and an exchange-correlation term
\beq
\overline{E}_{\rm xc}^\mu[n]=\overline{F}^\mu[n]-\overline{E}_{\rm H}^\mu[n]
\eeq
that needs to be approximated. The functional $\overline{E}_{\rm xc}^\mu[n]$
can, in turn, be divided into exchange and correlation in two different
ways. We can, in fact, define an exchange functional by using the Kohn-Sham
determinant $\Phi$,
\beq
\overline{E}_{\rm x}^\mu[n]  =  
\langle \Phi |\hat{W}_{ee}-\hat{W}^\mu|\Phi\rangle-
\overline{E}_{\rm H}^\mu[n],
\label{eq_Excomp}
\eeq
and then define the usual correlation energy functional 
$\overline{E}^\mu_c[n]$,
\beq
\overline{E}^\mu_c[n]=\overline{E}^\mu_{\rm xc}[n]-\overline{E}_{\rm x}^\mu[n],
\label{eq_Eccomp}
\eeq
but we can also define a multideterminantal (md) exchange 
functional~\cite{tca} by using the wavefunction $\Psi^\mu$,
\beq
\overline{E}_{\rm x, md}^\mu[n]  =  
\langle \Psi^\mu |\hat{W}_{ee}-\hat{W}^\mu|\Psi^\mu\rangle-
\overline{E}_{\rm H}^\mu[n],
\label{eq_Exmd}
\eeq
and then a corresponding correlation energy,
\beq
\overline{E}^\mu_{\rm c, md}[n]=\overline{E}^\mu_{\rm xc}[n]-\overline{E}_{\rm x, md}^\mu[n].
\label{eq_Ecmd}
\eeq
These two ways of splitting exchange and correlation play a role only
if we implement an optimized effective potential-like scheme, 
in analogy with what is usually done for the exact-exchange 
KS-DFT~\cite{OEP} [see Eq.~(\ref{eq_oepKS})].  
In the case of the multideterminantal exchange of Eq.~(\ref{eq_Exmd})
the problem can be reformulated as~\cite{tca}
\beq
E_0=\inf_{v^\mu}\left\{\langle\Psi^\mu_{v^\mu}| \hat{T}+\hat{W}_{ee}+\hat{V}_{ne}
|\Psi^\mu_{v^\mu}\rangle+\overline{E}^\mu_{\rm c, md}[n_{\Psi^\mu_{v^\mu}}]\right\},
\eeq
where $\Psi^\mu_{v^\mu}$ is obtained by solving the Schr\"odinger equation
corresponding to the hamiltonian $\hat{H}^\mu= \hat{T}+\hat{W}^\mu+\hat{V}^\mu$.
This equation is the generalization of Eq.~(\ref{eq_oepKS}) to the case of
the multideterminantal model system $\Psi^\mu$.

\subsection{Exact properties of the functionals}
\label{sec_prop}
From Eq.~(\ref{eq_FfromAC}) it is possible to derive exact properties 
of the functionals of Eqs.~(\ref{eq_Excomp})--(\ref{eq_Ecmd}) in the 
$\mu\to 0$ limit~\cite{julien,lsderf} and in the $\mu\to\infty$ 
limit~\cite{julien,GS3}.
\par
In the first case, all the functionals $\overline{E}_x^\mu$,
$\overline{E}_c^\mu$ [Eqs.~(\ref{eq_Excomp}) and~(\ref{eq_Eccomp})] and
$\overline{E}_{\rm x, md}^\mu$,
$\overline{E}_{\rm c, md}^\mu$ [Eqs.~(\ref{eq_Exmd}) and~(\ref{eq_Ecmd})]
tend to the KS functionals of Sec.~\ref{sec_KSDFT}. This limit
can be studied with perturbation theory: one finds that 
the way in which the functionals
approach the KS ones depends on wether the system is confined~\cite{julien} (atoms,
molecules) or extended~\cite{lsderf}.  
\par
The large-$\mu$ limit is the most interesting, since,
as shown by Eq.~(\ref{eq_FfromAC}), it always lies in the
range of $\mu$-values for which we want to construct approximations.
In this limit, all the functionals of Eqs.~(\ref{eq_Excomp})--(\ref{eq_Ecmd})
vanish and, since $\partial w^\mu(r_{12})/\partial \mu=\frac{2}{\sqrt{\pi}}e^{-\mu^2
r_{12}^2}$, Eq.~(\ref{eq_FfromAC}) shows that 
their large-$\mu$ behavior is determined by the short-range part (small $r_{12}$)
of  the spherically and system-averaged pair density $f^\mu(r_{12})$. 
It is possible to show~\cite{GS3} that, 
when $\mu\to\infty$, the small-$r_{12}$ part of $f^\mu(r_{12})$ is dominated by
\beq
f^\mu(r_{12})=f(0)\left[1+2\,r_{12}\, p_1(\mu r_{12})+\frac{2}{\sqrt{\pi}\mu}
\right],
\label{eq_expaf}
\eeq
where $f(0)$ is the ``on-top value'' (zero electron-electron distance) of
the physical (i.e., corresponding to the Coulomb interaction) $f(r_{12})$, and the
function $p_1(y)$ is equal to~\cite{GS3}
\beq
p_1(y) =  \frac{e^{-y^2}-2}{2\sqrt{\pi}\, y}
+\left(\frac{1}{2}+\frac{1}{4\,y^2}\right) \erf(y).
\label{eq_p1}
\eeq
Inserting Eq.~(\ref{eq_expaf}) (and, for the case of $E_x^\mu[n]$, the KS
$f^{\mu=0}(r_{12})$, corresponding to the KS determinant $\Phi$) 
into Eq.~(\ref{eq_FfromAC}) we find, for unpolarized systems~\cite{gill,julien,GS3}
\begin{eqnarray}
& & \overline{E}_x^{\mu\to\infty}[n]  =  -\frac{\pi}{4\mu^2}\int d\rv n(\rv)^2
+O\left(\frac{1}{\mu^4}\right),
\label{eq_largemuEx}         \\
& & \overline{E}_c^{\mu\to\infty}[n]  =  
\frac{\pi}{\mu^2}\left[f(0)-\frac{1}{4}\int d\rv n(\rv)^2\right]
+f(0)\frac{4\sqrt{2\pi}}{3\,
\mu^3}+O\left(\frac{1}{\mu^4}\right),          
\label{eq_largemuEc}         \\
& & \overline{E}_{\rm x, md}^{\mu\to\infty}[n]  =  
\frac{\pi}{\mu^2}\left[f(0)-\frac{1}{2}\int d\rv n(\rv)^2\right]
+f(0)\frac{4\sqrt{\pi}(2\sqrt{2}-1)}{3\,
\mu^3}+O\left(\frac{1}{\mu^4}\right),       
\label{eq_largemuExmd}         \\
& & \overline{E}_{\rm c, md}^{\mu\to\infty}[n]  =  
-f(0)\frac{4\sqrt{\pi}(\sqrt{2}-1)}{3\,
\mu^3}+O\left(\frac{1}{\mu^4}\right).
\label{eq_largemuEcmd}
\end{eqnarray}
These equations tell us that (i) if we use the definition of exchange with
the KS determinant, the functional $\overline{E}_x^{\mu}[n]$ is, 
for large $\mu$, 
exactly described by a local functional of the density; (ii) all the other
three functionals involve the physical ``on-top'' $f(0)$, which
is usually not available (its knowledge would require a very accurate calculation
for the physical system!). However, it is possible to construct approximations
for $f(0)$. For instance, for many systems the estimate of $f(0)$
from the local density approximation (i.e., by transfer from the uniform
electron gas model) is rather good~\cite{variesucorrSR}. Another possibility is
to use Eq.~(\ref{eq_expaf}) to estimate $f(0)$ from $f^\mu(0)$. This estimate
has the advantage of being without self-interaction error~\cite{GS3}. 

\subsection{Building approximate functionals}
\label{sec_approximations}
Following the same historical path of KS-DFT,
the simplest approximation one can think of is the local density (LDA),
\beq
\overline{E}_{\rm xc}^\mu[n]=\int d\rv\,n(\rv)\, [\epsilon_{\rm xc}(n(\rv))-
\epsilon_{\rm xc}^\mu(n(\rv))],
\eeq
where  $\epsilon_{\rm xc}(n)$ and $\epsilon_{\rm xc}^\mu(n)$ are
the exchange-correlation energy per particle of an electron gas of uniform
density $n$ with interaction $1/r_{12}$~\cite{cepald} and 
$w^\mu(r_{12})$~\cite{julIJQC,lsderf}, respectively. 
\par
Generalized-gradient approximations (GGA) for $\overline{E}_{\rm xc}^\mu$ 
have been also constructed \cite{julJCP,HSE,stoll,goll} along similar lines 
of KS-DFT~\cite{PBE}. In this context, we only mention a conceptually
simple approximation based on the exact properties of the previous 
Sec.~\ref{sec_prop}. It consists~\cite{julJCP} in a rational 
interpolation between
a given GGA KS functional at $\mu=0$, and zero at $\mu\to\infty$. The rational
interpolation is constrained to recover the exact large-$\mu$ behavior
of Eqs.~(\ref{eq_largemuEx})-(\ref{eq_largemuEc}), using the LDA approximation
for the physical on-top $f(0)$.
\par
All these ways of constructing approximate functionals are based
on the assumption of transferability of short-range
exchange and correlation effects from one system (the uniform 
electron gas) to the others. Another strategy that came out recently,
and that seems very well suited to describe short-range correlation
effects, consists in generating realistic APD $f^{\mu'}(r_{12})$ 
along the adiabatic connection [with $\mu'\ge\mu$, to be inserted
in Eq.~(\ref{eq_FfromAC})]
by solving
simple ``radial'' (unidimensional) equations~\cite{GS1,GS2} for a set of
``effective geminals'' $\psi^{\mu'}_i(r_{12})$,
\begin{eqnarray}
& & \left[-\frac{1}{r_{12}}\frac{d^2}{dr_{12}^2}r_{12}
+\frac{\ell (\ell+1)}{r_{12}^2}+w_{\rm eff}^{\mu'}(r_{12})\right] \psi^{\mu'}_i(r_{12})  =  \epsilon_i^{\mu'}\,
\psi_i^{\mu'}(r_{12}) 
\nonumber \\ 
& & \sum_i \vartheta_i|\psi_i^{\mu'}(r_{12})|^2  =  f^{\mu'}(r_{12}).
\label{eq_eff}
\end{eqnarray}
The effective interaction  $w_{\rm eff}^{\mu'}(r_{12})$ is (in principle)
the Lagrange multiplier
for the exact $f^{\mu'}(r_{12})$ (i.e., corresponding
to the wavefunction $\Psi^{\mu'}$ along the
adiabatic connection; in practice  $w_{\rm eff}^{\mu'}(r_{12})$
is approximated). In these equations the
rule for the occupancy $\vartheta_i$ 
of the effective geminals is chosen, for spin compensated systems, 
to be Slater-determinant-like:
occupancy 1 for even $\ell$ (singlet symmetry), occupancy
3 for odd $\ell$ (triplet symmetry), up to $N(N-1)/2$ occupied geminals.
This rule has been applied to solve the effective equations~(\ref{eq_eff})
in the uniform electron gas at $\mu=\infty$ (Coulombic interaction), 
with rather accurate 
results~\cite{GP1,DPAT1} when combined with simple approximations
for the effective interaction potential $w_{\rm eff}(r_{12})$.

\subsection{The calculation of $\Psi^\mu$: an illustrative example}
As said in the previous paragraphs, the model wavefunction 
$\Psi^\mu$ is, in most cases, computed with one of the traditional
wavefunction methods of quantum chemistry. We report here an illustrative
example of such calculation, and we then give in the next 
Sec.~\ref{sec_results} an
overview of recent results obtained using different methods for computing
$\Psi^\mu$, combined with different approximations for the functionals. 
\par
Since the only purpose of this paragraph is to investigate the wavefunction
part of the multideterminantal DFT, we only consider two very small systems,
the He and the Be atoms, for which it has been possible to construct very
accurate potentials $v^\mu$~\cite{julien}. In this way (i) we have essentially
no approximation on the functional part and we can  focus our attention
on the effect of approximations on the calculation of $\Psi^\mu$, and
(ii) we can produce accurate benchmark results to test our calculations.
Of course, this is not the general procedure in which we are finally interested.
The general procedure rather consists in using a given approximation (see
Sec.~\ref{sec_approximations}) for the
functional $\overline{E}^\mu_{\rm xc}[n]$ and the corresponding potential, and
combining it with  a wavefunction method to calculate $\Psi^\mu$, to finally
obtain the total ground state energy of the physical hamiltonian. 
This procedure has been followed with very promising results in the works
reviewed in the next Sec.~\ref{sec_results}.

Here, in order to illustrate
the efficiency of approximate wavefunction methods to treat $\Psi^\mu$, 
we thus proceed as follows~\cite{julien}.
We first construct, for each $\mu$, the model Hamiltonian $\hat{H}^{\mu}$ using an accurate potential $v^{\mu}$~\cite{julien} and compute accurately its ground-state energy, $E^{\mu} = \langle\Psi^\mu| \hat{H}^{\mu} |\Psi^\mu\rangle$, at the multi-reference configuration interaction with singles and doubles (MRCISD) level. We then compute various approximate ground-state energies, $E^{\mu}_{S} = \langle\Psi^\mu_S| \hat{H}^{\mu} |\Psi^\mu_S\rangle$, by using approximate configuration inteaction (CI) type wave functions $\Psi^{\mu}_{S}$ expanded into linear combinations of all the few Slater determinants generated from small orbital spaces $S$. The orbitals used are the natural orbitals of the Coulombic system calculated at the MRCISD level. The accuracy of the approximation for $\Psi^{\mu}_{S}$ can be assessed by looking at the difference between $E^{\mu}_{S}$ and $E^{\mu}$ 
\begin{equation}
\Delta E^{\mu}_{S}= E^{\mu}_{S} - E^{\mu}.
\end{equation}

\begin{figure}
\includegraphics[scale=0.75]{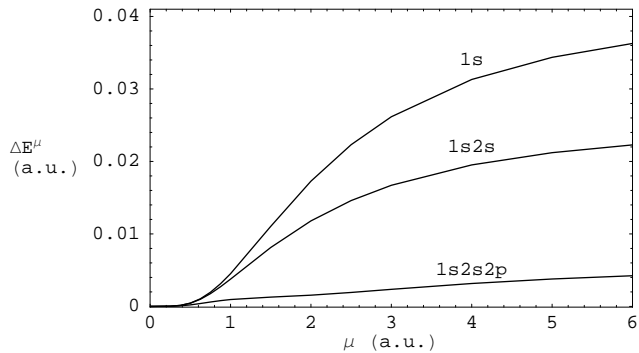}
\caption{Ground-state energy differences $\Delta E^{\mu}_{S}= \langle\Psi^{\mu}_{S} |\hat{H}^{\mu}| \Psi^{\mu}_{S}\rangle - \langle\Psi^{\mu} |\hat{H}^{\mu}| \Psi^{\mu}\rangle$ where $\Psi^\mu$ is an accurate wave function and $\Psi^{\mu}_{S}$ are approximate wave functions generated from small orbital spaces $S=1s$, $S=1s2s$ and $S=1s2s2p$, as a function of $\mu$ for the He atom.
}
\label{fig_CI_He}
\end{figure}

The differences $\Delta E^{\mu}_{S}$ are plotted as a function of $\mu$ in Fig.~\ref{fig_CI_He} for the He atom with the orbital spaces $S=1s$, $S=1s2s$ and $S=1s2s2p$. One sees that, in the Coulombic limit, $\mu \to \infty$, the reduction of the orbital space leads to important errors in the energy. When $\mu$ is decreased, i.e. when the interaction is reduced, the errors due to limited orbital spaces get smaller and smaller. For instance, at $\mu=1$, using only the single-determinant wave function $\Psi^\mu_{1s}$, leads to an error $\Delta E^{\mu}_{1s}$ of less than $0.005$ Hartree. 

\begin{figure}
\includegraphics[scale=0.75]{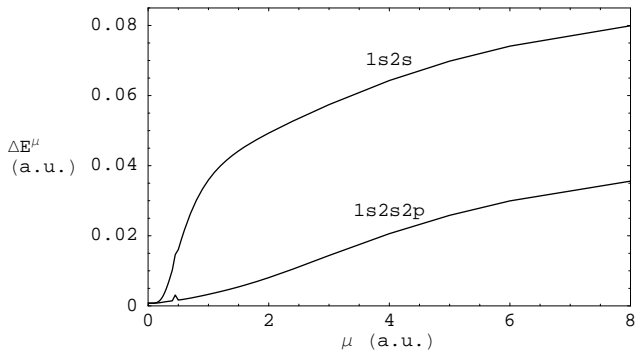}
\caption{
Ground-state energy differences $\Delta E^{\mu}_{S}= \langle\Psi^{\mu}_{S} |\hat{H}^{\mu}| \Psi^{\mu}_{S}\rangle - \langle\Psi^{\mu} |\hat{H}^{\mu}| \Psi^{\mu}\rangle$ where $\Psi^\mu$ is an accurate wave function and $\Psi^{\mu}_{S}$ are approximate wave functions generated from small orbital spaces $S=1s2s$ and $S=1s2s2p$, as a function of $\mu$ for the Be atom.
}
\label{fig_CI_Be}
\end{figure}

The case of the Be atom with the orbital spaces $S=1s2s$ and $S=1s2s2p$ is reported in Fig.~\ref{fig_CI_Be}. Because of the near-degeneracy of the $2s$ and $2p$ levels, the inclusion of $2p$ configurations in the wave function is important, quite independently of the electron-electron interaction. Indeed, the difference $E^{\mu}_{1s2s} - E^{\mu}_{1s2s2p}$  remains large for almost all $\mu$'s. On the contrary, the error of the calculation where the $2p$ orbitals are included, $\Delta E^{\mu}_{1s2s2p}$, quickly falls off when $\mu$ is decreased. Again, for $\mu=1$ for instance, the error $\Delta E^{\mu}_{1s2s2p}$ given by the few-determinant CI-type wave function $\Psi^\mu_{1s2s2p}$ is less than $0.005$ Hartree. 

Therefore, the modification of the interaction enables to increase the accuracy of CI-type wave function calculations, or equivalently for a fixed target accuracy, decrease the effort of the calculation by reducing the orbital space. The crucial point for this effect to appear seems to  be the reduction of the electron-electron interaction compared to the Coulomb interaction rather than the long-range character of the modified interaction.

\section{Some recent results: a short overview}
\label{sec_results}
The idea of using the partially interacting model system of Eq.~(\ref{eq_Fmu}) 
has been explored with different techniques to compute the
wavefunction $\Psi^\mu$, and using different approximations for the 
functionals of Sec.~\ref{sec_functionals}. We review very briefly some
of the corresponding results.
\par 
The LDA functional for $\overline{E}_{\rm xc}^\mu[n]$ has been combined
with configuration interaction (CI) to handle the wavefunction $\Psi^\mu$
in Refs.~\cite{sav1,sav2}; the method has been applied to atoms and small molecules.
Again using LDA for the short-range functional, 
in Ref.~\cite{janos} it has been shown that 
long-range van der Waals forces in rare gas dimers can be well described by
using second-order perturbation theory for $\Psi^\mu$.
\par
A short-range GGA functional has been combined with the coupled-cluster 
(CC) method
to describe $\Psi^\mu$, with very good results for small molecules,
both for the closed and the open shell cases~\cite{stoll,goll}: in particular,
the results from multideterminantal DFT for small and medium basis-set
sizes are better of both the pure DFT result and the pure CC result. In
Ref.~\cite{JulTesi}, a different GGA functional~\cite{HSE,julien} 
has been used, 
and $\Psi^\mu$ has been determined by multi-configuration self
consitent field (MCSCF). The corresponding application to systems
in which near-degeneracy effects play a major role is promising, although
self-interaction errors in the functionals are still a problem in some 
cases~\cite{JulTesi}.
\par
Realistic $f^{\mu'}(r_{12})$ from Eqs.~(\ref{eq_eff})
have been generated for two-electron atoms in
Ref.~\cite{GS1}: with a very simple approximation for 
$w_{\rm eff}^{\mu'}(r_{12})$ (inspired to the one 
used for the electron gas at $\mu=\infty$), extremely accurate 
$\overline{E}_{\rm xc}^\mu[n]$ have been obtained. This new strategy
of producing functionals is still at a very early stage, and several
steps are needed for its development. In particular, we need
better approximations for $w_{\rm eff}^{\mu'}(r_{12})$, and  the implementation
of an efficient algorithm to couple the radial equations (\ref{eq_eff})
with the method used to solve the model hamiltonian $\hat{H}^\mu$.
\par
In conclusions,
changing the model Hamiltonian in DFT calculations
seems to be a promising alternative  to the construction of better 
exchange-correlation functionals for standard Kohn-Sham DFT. Of course,
the method has been under development only in the last decade, and it 
still needs
further improvement and investigation in many aspects.

\bibliographystyle{amsalpha}

\begin{thebibliography}{A}

\bibitem{pople} see, e.g.,
       J.A. Pople, \textit{Nobel Lecture: Quantum chemical models},
       Rev. Mod. Phys. (1999) \textbf{71}, 1267--1274.
\bibitem{mitas} see, e.g.,
 W.M.C. Foulkes, L. Mitas, R.J. Needs, and G. Rajagopal,
\textit{Quantum Monte Carlo simulations of solids},
       Rev. Mod. Phys. (2001) \textbf{73}, 33--83.
\bibitem{kohnnobel} see, e.g.,
W. Kohn, \textit{Nobel Lecture: Electronic structure of matter -- 
wave functions and density functionals},
Rev. Mod. Phys.  (1999) \textbf{71}, 1253--1266.
\bibitem{coleman} see, e.g.,
A.J. Coleman and V.I. Yukalov, \textit{Reduced density matrices :
Coulson's challenge}, Springer, Berlin, 2000.
\bibitem{HK} P. Hohenberg and W. Kohn, \textit{Inhomogeneous Electron Gas}
Phys. Rev. (1964) \textbf{136}, B864-B871. 
\bibitem{mel} M. Levy, 
\textit{Universal variational functionals of electron densities, 1st-order
reduced density-matrices, and natural spin-orbitals and solution of the
v-representability problem},
Proc. Natl. Acad. Sci. U.S.A. (1979) {\bf 76}, 
6062-6065.
\bibitem{lieb} E.H. Lieb, Int. J. Quantum Chem.,
\textit{Density functionals for Coulomb-systems}
(1983) {\bf 24}, 243-277.
\bibitem{KS}
W. Kohn and L.J. Sham, \textit{Self-Consistent Equations Including
Exchange and Correlation Effects},
Phys. Rev. (1965) \textbf{140}, A1133-A1138.
\bibitem{science}
see, e.g., A. E. Mattsson, \textit{In pursuit of the "divine" functional},
Science (2002), \textbf{298}, 759-760.
\bibitem{OEP} see, e.g.,
S. K\"ummel and J.P. Perdew, 
\textit{Optimized effective potential made simple: Orbital functionals, orbital shifts, and the exact Kohn-Sham exchange potential},
Phys. Rev. B (2003) {\bf 68}, 035103 1-15;
W. Yang and Q. Wu, 
\textit{Direct Method for Optimized Effective Potentials in 
Density-Functional Theory},
Phys. Rev. Lett. {\bf 89}, 143002 1-4 (2002), and references therein.
\bibitem{adiabatic}
see, e.g., A. Savin, F. Colonna, R. Pollet,
\textit{Adiabatic connection approach to density functional theory of electronic systems},
Int. J. Quantum Chem.  (2003) \textbf{94}, 166-190, and references therein.
\bibitem{weitao}
W. Yang, 
\textit{Generalized adiabatic connection in density functional theory},
J. Chem. Phys.  (1998), {\bf 109}, 10107-10110.
\bibitem{variesucorrSR} 
see, e.g., K. Burke, J. P. Perdew, and M. Ernzerhof,
\textit{Why semilocal functionals work: Accuracy of the on-top pair density and importance of system averaging},
J. Chem. Phys. (1998) \textbf{109}, 3760-3771.  
\bibitem{rey} J. Rey and A. Savin, \textit{Virtual space level shifting and correlation energies},
Int. J. Quantum Chem. (1998), \textbf{69}, 581-590.
\bibitem{claudinetesi} C. Gutl\'e, \textit{Espace orbitalaires et th\'eorie de la fonctionelle
de la densit\'e}, PhD thesis, Universit\'e Pierre et Marie Curie (Paris, France). Full text
(in French) available at {\tt http://www.lct.jussieu.fr/DFT/}
\bibitem{pickett} G.E. Engel, and W.E. Pickett, 
\textit{Investigation of density functionals to predict both ground-state properties and band structures},
Phys. Rev. B  (1996) \textbf{54}, 8420-8429.
\bibitem{tca} J. Toulouse, P. Gori-Giorgi and A. Savin,
\textit{A short-range correlation energy density 
functional with multi-determinantal
reference}, Theor. Chem. Acc. (2005), \textbf{114}, 305-308.
\bibitem{julien} J. Toulouse, F. Colonna, A. Savin,
\textit{Long-range/short-range separation of the electron-electron interaction in density functional theory},
Phys. Rev. A (2004) \textbf{70}, 062505 1-16.
\bibitem{lsderf} S. Paziani, S. Moroni, P. Gori-Giorgi, and G.B. Bachelet,
\textit{Local-spin-density functional for multideterminant density functional theory}, Phys. Rev. B  (2006) \textbf{73}, 155111 1-9.
\bibitem{GS3}  P. Gori-Giorgi and A. Savin,
\textit{Properties of short-range and long-range correlation energy density functionals from electron-electron coalescence}, Phys. Rev. A (2006) \textbf{73}, 
032506 1-9.
\bibitem{gill} P.M.W. Gill and R.D. Adamson, 
\textit{A family of attenuated Coulomb operators},
Chem. Phys. Lett. (1996) \textbf{261}, 105-110.
\bibitem{cepald} 
D. M. Ceperley and B.J. Alder, \textit{Ground State of the Electron Gas by a Stochastic Method}, Phys. Rev. Lett.  (1980) \textbf{45}, 566-569.
\bibitem{julIJQC} J. Toulouse, A. Savin, and H.-J. Flad, 
\textit{Short-range exchange-correlation energy of a uniform electron gas with modified electron-electron interaction},  Int. J. Quantum. Chem.  (2004) \textbf{100}, 1047.
\bibitem{julJCP} 
 J. Toulouse, F. Colonna, and A. Savin, \textit{Short-range exchange and correlation energy density functionals: beyond the local density approximation}, 
J. Chem. Phys. (2005) \textbf{122}, 014110 1-10.
\bibitem{HSE}
J. Heyd, G.E. Scuseria, and M. Ernzerhof
 \textit{Hybrid functionals based on a screened Coulomb potential}, J. Chem. Phys.
(2003), \textbf{118} 8207-8215.
\bibitem{stoll} E. Goll, H.-J. Werner, and H. Stoll, \textit{A short-range
gradient-corrected density functional in long-range coupled-cluster calculations
for rare gas dimers}, Phys. Chem. Chem. Phys. (2005) \textbf{7}, 3917-3923.
\bibitem{goll}
E. Goll, H.-J. Werner, H. Stoll, T. Leininger, P. Gori-Giorgi, and A. Savin,
\textit{A short-range gradient-corrected spin density functional in combination with long-range coupled-cluster methods: Application to alkali-metal rare-gas dimer}, Chem. Phys., in press.
\bibitem{PBE} 
J. P. Perdew, K. Burke, and M. Ernzerhof, \textit{Generalized Gradient Approximation 
Made Simple}, Phys. Rev. Lett.  (1996) \textbf{77}, 3865-3868.
\bibitem{GS1}
P. Gori-Giorgi and A. Savin,
\textit{Simple model for the spherically- and system-averaged pair density: Results for two-electron atoms},  
Phys. Rev. A  (2005) \textbf{71}, 032513 1-9.
\bibitem{GS2}
P. Gori-Giorgi and A. Savin,
\textit{System-adapted correlation energy density functionals from effective pair interactions}, Phil. Mag. (2006) \textbf{86}, 2643-2659.
\bibitem{GP1}
P. Gori-Giorgi and J.P. Perdew,
\textit{Short-range correlation in the uniform electron gas: Extended Overhauser model},  Phys. Rev. B  (2001) \textbf{64}, 155102 1-8.
\bibitem{DPAT1}
B. Davoudi, M. Polini, R. Asgari, and M. P. Tosi,
\textit{Self-consistent Overhauser model for the pair distribution function of an electron gas in dimensionalities D = 3 and D = 2},
 Phys. Rev. B  (2002) \textbf{66}, 075110 1-8. 
\bibitem{sav1}
T. Leininger, H. Stoll, H.-J. Werner, and A. Savin, 
\textit{Combining long-range configuration interaction with short-range density functionals},
Chem. Phys. Lett. (1997)
\textbf{275}, 151-160.
\bibitem{sav2}
R. Pollet, A. Savin, T. Leininger, and H. Stoll,
\textit{Combining multideterminantal wave functions with density functionals to handle near-degeneracy in atoms and molecules}, 
J. Chem. Phys. (2002) \textbf{116}, 1250-1258.
\bibitem{janos}
J. G. Angyan, I. C. Gerber, A. Savin, and J. Toulouse, \textit{van der Waals forces in density functional theory: perturbational long-range electron interaction corrections}, Phys. Rev. A  (2005) \textbf{72}, 012510 1-9.
\bibitem{JulTesi} J. Toulouse, \textit{Extension multid\'eterminantale de la m\'ethode de Kohn-Sham en th\'eorie de la fonctionnelle
de la densit\'e par d\'ecomposition de l'interaction \'electronique 
en contributions de longue port\'ee et de courte port\'ee}, PhD thesis, Universit\'e Pierre et Marie Curie (Paris, France). Full text
(in French) available at {\tt http://www.lct.jussieu.fr/toulouse/}.
\end{thebibliography}

\end{document}